\documentstyle[preprint,aps]{revtex}

\newcommand{\es}     {\epsilon}

\newcommand{\lm}     {\lambda}
\newcommand{\M}     {{\mathcal M}}
\newcommand{\no}     {\nonumber}
\begin{document}
\draft
\preprint{
\vbox{\hbox{KIAS-P98049}
      \hbox{SNUTP\hspace*{.2em}98-143}}
}
\title{
Four Light Neutrinos in Singular Seesaw Mechanism 
\\
with Abelian Flavor Symmetry
}
\author{
Chun Liu$^{\:a}$ and Jeonghyeon Song$^{\:b}$
}
\vspace{1.5cm}
\address{
$^a$Korea Institute for Advanced Study, 207-43 Chungryangri-dong,
Dongdaemun-ku,\\ 
Seoul 130-012, Korea\\
$^b$Center for Theoretical Physics, Seoul National University,\\
Seoul 151-742, Korea
}
\maketitle
\thispagestyle{empty}
\setcounter{page}{1}
\begin{abstract}
The four light neutrino scenario, which explains the atmosphere, solar 
and LSND neutrino experiments, is studied in the framework of the seesaw
mechanism.  By taking both the Dirac and Majorana mass matrix of neutrinos 
to be singular, the four neutrino mass spectrum consisting of two almost 
degenerate pairs separated by a mass gap $\sim 1$~eV is naturally generated.
Moreover the right-handed neutrino Majorana mass can be at $\sim 10^{14}$~GeV 
scale unlike in the usual singular seesaw mechanism. Abelian flavor symmetry 
is used to produce the required neutrino mass pattern. A specific example 
of the flavor charge assignment is provided to show that maximal mixings 
between the $\nu_\mu-\nu_\tau$ and $\nu_e-\nu_s$ are respectively attributed 
to the atmosphere and solar neutrino anomalies while small mixing between 
two pairs to the LSND results.  The implication in the other fermion masses 
is also discussed.
\end{abstract}


\newpage

\section{Introduction}
\label{sec:introduction}

The recent Super-Kamiokande data on the zenith-angle-dependent
deficit of atmospheric $\mu-$type neutrinos 
have provided compelling evidence of
neutrino masses and mixing\cite{superk}.  It might be
the first discovery of physics beyond the standard model (SM), and has 
drawn a lot of theoretical attentions.  
The mass pattern of neutrinos is valuable information for the
exploration of the physics related to the flavor puzzle in the SM.  
Theoretically, several mechanisms have been suggested
to accommodate massive neutrinos.  
One of the most popular scenarios is the seesaw mechanism
which naturally explains the smallness of neutrino mass
by introducing heavy right-handed neutrinos \cite{seesaw}.  
Another example is the supersymmetric extension of the SM 
with R-parity violation.
The trilinear lepton number violating interactions induce small neutrino
masses at the loop level \cite{rparity}.  
Since the whole flavor problem in the SM has not been understood yet,
the detailed neutrino mass pattern in these models remains indefinite.

The observations of the atmospheric and solar neutrinos 
give the information about the neutrino mass squared 
differences and mixing angles under the assumption of neutrino oscillations.
According to the Super-Kamiokande data,
about 35\% of the $\mu-$type atmospheric neutrinos change
their flavor into non-$e-$type neutrinos, implying
$\Delta m_{\mu x}^2\simeq 2.2 \times 10^{-3}$ eV$^2$
and $\sin^2 2 \theta_{\mu x} \simeq 1\, (x \neq e)$ \cite{superk}.
The solar neutrino deficit problem \cite{Bahcall} can be explained 
by either the matter-enhanced oscillation (MSW effects) \cite{msw}
or the vacuum oscillation\cite{vac}.
The MSW solution allows two parameter spaces:
$\Delta m_{ey}^2\simeq 5 \times 10^{-6}$ eV$^2$ with
$\sin^2 2 \theta_{e y} \simeq 6 \times 10^{-3}$, and
$\Delta m_{ey}^2\simeq 2 \times 10^{-5}$ eV$^2$ with
$\sin^2 2 \theta_{e y} \simeq 0.8$.
The vacuum oscillation solution is
$\Delta m_{ey}^2\simeq 8 \times 10^{-11}$ eV$^2$ with
$\sin^2 2 \theta_{e y} \simeq 0.8$. 
If only the solar and atmospheric neutrino data are taken
into account,
they can be understood in the frameworks with three light neutrinos 
\cite{three,three2}.  

If we further consider the LSND experiment \cite {lsnd}, however,
something previously unexpected has to be introduced in the theory.
When ascribing the detection of flavor-changing events to
neutrino oscillations, 
the experiment indicates that $\Delta m_{e\mu}^2\simeq 1$ eV$^2$ and
$\sin^2(2\theta_{e\mu})\simeq 10^{-2}$.  To accommodate all the above
experiments, three light neutrinos are not enough, at least one
additional light neutrino $\nu_s$ is needed \cite{four}.  It should be
sterile from the $Z^0$ decay.  Even though the KARMEN 
\cite{karmen} group has recently reported that 
large part of the favored parameter region of LSND is excluded,
full confirmation of the LSND results still awaits future
experiments.
For example, a particular value of
$\Delta m_{e\mu}^2 \sim 6$ eV$^2$ 
compatible with the LSND results does not contradict the KARMEN data
since this $\Delta m^2$ is examined most sensitively at LSND
while least sensitively at KARMEN.
Moreover, neutrinos with mass scale of eV play an important role 
in understanding the dark matter problem.
In astrophysics, it has been known that the Cold + Hot Dark Matter
cosmological models (CHDM) agree best with the data 
on the cosmic microwave background anisotropies and the
large-scale distribution of galaxies and clusters 
in the nearby universe \cite{cos}.
With about 70\% cold dark matter and about 10\% 
baryonic matter, few-eV neutrino mass is requisite to
account for the rest 20\% hot dark matter.

In this paper, the four light neutrino scenario is adopted.
Phenomenological studies \cite{four} have shown that the following mass
patterns are favored.  In terms
of the mass eigenstates, the four neutrinos are grouped into two pairs
which are separated by a gap of $\sim 1$ eV.  The two neutrinos in a
pair are almost degenerate compared to the gap.  The atmosphere
neutrino anomaly can be explained by any of two pairs, and the solar
neutrino deficit by the other.  The LSND data is explained by two
neutrinos in different pairs.  In terms of weak eigenstates,
approximately speaking, $\nu_{\tau}$ can pair with either
$\nu_{\mu}$ or $\nu_e$.

How to construct this spectrum of very small masses
of four neutrinos is a theoretically
challenging problem\cite{gen_four}.
One of the most appealing explanation for the smallness of neutrino masses
is the seesaw mechanism which introduces three very heavy right-handed
neutrinos $N_{\alpha (=e, \mu, \tau)}$ of Majorana masses $\sim M$.
Moreover the generic presence of the $N_{\alpha (=e, \mu, \tau)}$'s
in many extensions of the SM like SO(10) GUT's and E$_6$ string 
theories adds more charms to the mechanism.
Since the ordinary seesaw mechanism predicts
three light neutrinos and three very heavy neutrinos,
however, there is no room for light sterile neutrinos.
Even though the so-called singular seesaw mechanism,
where the Majorana mass matrix of the $N_{\alpha (=e, \mu, \tau)}$'s
is singular, has been suggested for this problem,
a drawback occurs such that the characteristic mass scale 
of lepton number violation is too low.
It deserts one of the strongest merits of the seesaw
mechanism itself.

In this letter we propose a four neutrino scenario in the 
framework of the seesaw mechanism,
which maintains the $M$ at GUT scales
so to appreciate all the original attractions of the seesaw mechanism,
and naturally produces the mass spectrum such that two
almost degenerate pairs are separated.
In addition,
the physical origin of our scenario in the view point
of the Abelian flavor symmetry is also discussed
through a specific example to
explain all the three neutrino anomalies.

\section{A Model for Four Light Neutrinos}
\label{sec:fs}

Natural generation of neutrino masses much lighter than the electroweak scale
brings popularity to the seesaw mechanism.
As introducing three right-handed neutrinos $N_{\alpha (=e, \mu, \tau)}$
with Majorana mass $\sim M$,
ordinary seesaw mechanism have two mass scales, 
the heavy neutrino mass $M$ and
the seesaw suppressed neutrino mass $m^2/M$,
where the $m$ is the Dirac mass.
Assigning the $m$ at the electroweak scale, 
the SM singlets $N_{\alpha (=e, \mu, \tau)}$'s can have masses in the
phenomenologically interesting range such that
$M\sim 10^{13}-10^{16}$~GeV corresponding to
a light neutrino mass $m_\nu \sim 10^{-3} -1$~eV.
In order to accommodate three neutrino experiment results
one of the right-handed neutrinos should be light.
This requires the singular seesaw mechanism:
the right-handed neutrino Majorana mass matrix
is singular\cite{ssm1}.
In Ref. \cite{ssm2}, this mechanism was used to produce the four light
neutrino mass pattern.
In their approach, however, a drawback appears such that
the mass of the heavy right-handed neutrinos is at keV scale,
which is further explained by introducing double seesaw mechanism.
This is because in Ref. \cite{ssm2} three mass scales, $M$, $m$, and
$m^2/M$ are introduced.
When the $m$ is charged with the LSND results and
hot dark matter while the $m^2/M$ with the solar neutrino
problem, the $M$ becomes much smaller than the GUT scale.

We observe that
if the Dirac mass matrix
is also singular, the right-handed neutrino
mass can be pulled back to GUT scale or so
even in singular seesaw mechanism.
For illustrative purpose, the mass matrix of the
following simple form is considered:
\begin{equation}
\label{1}
{\cal M} = \left[\begin{array}{cccccc}
0 & 0      & 0      & 0 & 0      & 0      \\
0 & 0      & 0      & 0 & m_{22} & m_{23} \\
0 & 0      & 0      & 0 & m_{32} & m_{33} \\
0 & 0      & 0      & 0 & 0      & 0      \\
0 & m_{22} & m_{32} & 0 & M_{22} & M_{23} \\
0 & m_{23} & m_{33} & 0 & M_{23} & M_{33} \\
\end{array}
\right]
\,,
\end{equation}
which is of rank four.
In the mass spectrum, there are two heavy neutrinos
of masses $\sim M$, two light neutrinos of masses $\sim m^2/M$, and two
massless
neutrinos.  In other words, the three scales in this case are $M$,
$m^2/M$, and $0$.
The four light neutrinos are naturally divided into
two pairs with a mass hierarchy
of $\sim m^2/M$, and each pair consists of two
degenerate mass eigenstates.

Thus the seesaw mechanism,
under the assumption that both the Dirac and Majorana mass
matrices are singular,
can naturally produce the looking-bizarre 
but required mass pattern to explain the three neutrino
anomalies.
Moreover, the $M$ can be around $10^{13}$ GeV
if the $m$ is taken to be at the electroweak scale and
the $m^2/M$ at $1$ eV scale,
maintaining a merit of the seesaw mechanism such that it can be easily
implemented in numerous theories for physics beyond the SM.

Now let us explore the physical origin 
of the singular seesaw mechanism,
which could lead to the specific neutrino mass matrix texture
in (\ref{1}).
It is natural to expect that there exists some symmetry 
to induce such neutrino mass pattern.  
This symmetry should also provide the large mixing for the 
atmosphere neutrino anomaly, which is no longer automatic
in (\ref{1})\footnote{Compared to the neutrino mass matrix
which yields a degenerate neutrino pair with maximum mixing
in the original singular seesaw mechanism\cite{ssm2},
(\ref{1}) gives vanishing mass to the two neutrinos.
In our case, their mass eigenstates can be always rotated
to the weak eigenstates.}.
Furthermore, a softly breaking of the symmetry 
is necessary to generate small masses for two massless neutrinos
and to lift the degeneracy in each pair.

It is known that Abelian flavor symmetry breaking by small parameters
could be an answer of the hierarchy in the fermion masses\cite{fn}.
We will use it to discuss the neutrino masses.  
In the following discussion, supersymmetry is implied.  The flavor symmetry is 
spontaneously broken by a vacuum expectation value (VEV) of an electroweak 
singlet field $X$.
As long as the flavor charges balance under the Abelian flavor symmetry,
the following interactions are allowed:
\begin{equation}
\label{2}
L_{\alpha} H N_{\beta} \left(
\frac{ \langle X \rangle_{\rm VEV} } {\Lambda}
\right)^{ m_{\alpha\beta}}
\,,
\quad
M N_\alpha N_\beta 
\left(
\frac{ \langle X \rangle_{\rm VEV} } {\Lambda}
\right)^{ n_{\alpha\beta}}
\,,
\end{equation}
where $L_\alpha$ ($\alpha=e$, $\mu$, $\tau$), $H$,
and  $N_\alpha$  denote the lepton 
doublets, one Higgs field, and the right-handed neutrino fields,
respectively.  
The $\Lambda$ is the flavor 
symmetry breaking scale, and the condition $m_{\alpha\beta}$, 
$ n_{\alpha\beta} \geq0$ is required for
the holomorphy of the superpotential.  
The order parameter for 
this new symmetry is defined by
\begin{equation}
\label{3}
\lm \equiv \frac{ \langle X \rangle_{\rm VEV} } {\Lambda}
\ll 1
\,.
\end{equation}
We intend the above neutrino mass pattern in (\ref{1})
to be achieved by the selection rules for (\ref{2}) 
through proper assignment of the flavor charges 
to $L_\alpha$ and $N_\alpha$.  
A further requirement is that the atmosphere
neutrino anomaly is due to the $\nu_\mu-\nu_\tau$ oscillation.  
Compared to analogous analysis for three light neutrino 
scenario which does not count the LSND result \cite{three2}, 
the choice of the flavor charges here is 
more limited.  One tricky point is that one of the right-handed neutrino
masses is made to be vanishingly small.  

As a specific example, we consider
the following assignment of the Abelian flavor charges:
\begin{eqnarray}
\label{assignment}
&& L_e(2t-a),~~L_\mu(a),~~L_\tau(-a-2),~~
\\ \no
&&
E^c_e(-x),~~E^c_{\mu}(-a+6),~~E^c_{\tau}(a+6),
\\ \no
&& N_e(2r+a),~~N_\mu(-a),
~~N_\tau(a+2),~~X(-2),
\end{eqnarray}
where the integers $a$, $t$, and $r$ are constrained as
\begin{equation}
\label{parameter-space}
1<t+1 < a<x<r
\,.
\end{equation}
The $E^c_{\alpha}$'s are the anti-particle fields of the SU(2) singlet
charged leptons.  In order to obtain the physical mixing angles of neutrinos,
we should simultaneously take into account of the mass matrix for the charged
lepton sector.
The fields of gauge bosons and Higgs' possess
vanishing flavor charges.
It is to be noted that all the flavor charges
for the second and third generations
are expressed by a single parameter $a$.

The flavor charge assignment in (\ref{assignment}) and 
(\ref{parameter-space}) produces the Dirac and Majorana mass 
matrices of neutrinos as
\begin{equation}
\label{m-nu}
\M_D = m
\left[
\begin{array}{ccc}
Y_{11} \lm^{r+t} & 0 & Y_{13} \lm^{t+1} \\
Y_{21} \lm^{r+a} & 1 & Y_{23} \lm^{a+1} \\
Y_{31} \lm^{ r-1} & 0 &1
\end{array}
\right]
\,, \quad
\M_M = \lm M
\left[
\begin{array}{ccc}
\zeta_1 \lm^{2r+a-1} & \zeta_2 \lm^{ r-1}  & \zeta_3 \lm^{r+a}\\
\zeta_2 \lm^{ r-1}   & 0 & 1 \\
\zeta_3 \lm^{r+a}   &   1   & \zeta_4 \lm^{a+1}
\end{array}
\right]
\,,
\end{equation}
and the mass matrix of the charged leptons as
\begin{equation}
\label{m-l}
\M_l =
m \lm^2
\left[
\begin{array}{ccc}
0 	&	0	&	\eta_{13} \lm^{t+1} \\
0	&	\eta_{22}\lm	&	\eta_{23} \lm^{a+1} \\
0	&	0	&	1
\end{array}
\right]
\,,
\end{equation}
where $Y$'s, $\zeta$'s and $\eta$'s are order one coefficients.  
To the leading order,
only tau lepton acquires mass $\lm^2 m$ while the muon and the
electron remain massless.
The mass matrix of four light neutrinos is obtained as follows, 
to the leading order,
\begin{equation}
\label{M0}
\M_{\nu}^{(0)}
\simeq
\frac{\es^2 M}{\lm}
\left[
\begin{array}{cccc}
0 & 0 & 0 & 0 \\
0 & 0 & -1 & 0 \\
0 & -1 & 0 & 0 \\
0 & 0 & 0 & 0 
\end{array}
\right]
\,.
\end{equation}
Here the $\es$ denotes the ratio of the weak scale to the GUT scale,
$\es \equiv m/M$.
The neutrino mass spectrum due to $\M_{\nu}^{(0)}$ is
\begin{equation}
\M_{\nu}^{(0)}
\Longrightarrow
\left[~~
m_{\nu_1}=m_{\nu_2}=0,\quad m_{\nu_3}=m_{\nu_4}=\frac{\es^2 M}{\lm},\quad
\sin\theta_{34}=\frac{1}{\sqrt{2}}
~~\right]
\,.
\end{equation}
To attribute the LSND data to the oscillation between two groups,
we require
\begin{equation}
\label{LSND}
\frac{m^2 }{\lm M} \sim 1 \, {\rm eV}
\,,
\end{equation}
which can be satisfied with the following masses:
\begin{equation}
\label{LSND-parameter}
m \simeq 10^2 \, {\rm GeV}\,, \quad
\lm M \sim 10^{13}  \, {\rm GeV}\,.
\end{equation}
One of the merits of our charge assignment is that we can obtain
larger mass scale for $M$, according to the small value $\lm$.
If $\lm \sim 10^{-1}$,
which is a typical order of Cabbibo angle,
the tau lepton mass is properly obtained:
\begin{equation}
m_\tau \sim  \lm^2 m \sim 1\,{\rm GeV}~.
\end{equation}

The full mass matrix of the charged leptons in (\ref{m-l}) 
is solved by the standard method.  
The eigenvalues of the $\M_l$ are $0$, $ \lm^3 m$, and $ \lm^2 m$.
In our mechanism with the flavor charge assignment
in (\ref{assignment}) and (\ref{parameter-space}),
the muon acquires the mass with appropriate 
order of magnitude ($\sim$100 MeV) but the
electron is left massless
even under the Abelian flavor symmetry breaking.
The $\M_l $ is diagonalized by
\begin{equation}
R_l^L \M_l \; R_l^{R\dagger} = {\rm Diag}\;  ( m_e, m_\mu, m_\tau)
\,.
\end{equation}
Since the $ R_l^L $ diagonalizes the hermitian mass-squared matrix
$\M_l \M_l^\dagger $,
we have
\begin{equation}
R_l^L =
\left[
\begin{array}{ccc}
1	&	0	&	0 \\
0	&	1	&	-\eta_{23} \lm^{a+1} \\
0	&	\eta_{23} \lm^{a+1}  & 1
\end{array}
\right]
\,.
\end{equation}
In the neutrino sector the mass matrix of four light neutrinos
can be obtained by the method described in Ref. \cite{ssm2}.
The $\M_M$ is diagonalized to give three eigenvalues
$\sim  \lm^{ {2r+a} } M$, $M$, and $-M$
by a rotation matrix $R_M$,
\begin{equation}
\label{15}
R_M=
\left[
\begin{array}{ccc}
1   & 0 & -  \zeta_2 \lm^{ r-1}\\
\zeta_2 \lm^{ r-1} /\sqrt{2 }
&	{1}/{\sqrt{2}}	& 	 {1}/{\sqrt{2}} \\
\zeta_2 \lm^{ r-1} /\sqrt{2 }
&	-{1}/{\sqrt{2}}	& 	 {1}/{\sqrt{2}} 
\end{array}
\right]
\,.
\end{equation}
We finally have the symmetric mass matrix of four light neutrinos,
\begin{equation}
\label{16}
\M_{\nu}=
\frac{\es^2 M}{\lm}
\left[
\begin{array}{cccc}
0	& \lm^{t+1}	&  0 	& \lm^{r+t+1}/\es  \\
\lm^{t+1}  &   \lm^{a}      &    -1    	&   {\lm^{r+a+1}}/{\es} \\
0      &  -1           &  \lm^{a+1}  		& \lm^r/\es   \\
\lm^{r+t+1}/\es  &   \lm^{r+a+1}/\es  &  \lm^r/\es & \lm^{2r+a+1}/\es^2    
\end{array}
\right]
\,,
\end{equation}
where the charged lepton mixing effects are incorporated so that
the charged lepton fields have been rotated to be mass eigenstates.
In (\ref{16}) each element denotes the order of magnitude estimate.
Now the mixing angle for the LSND results comes from
the $\lm^r/\es$ term, as
\begin{equation}
\label{LSND-mixing}
\sin \theta_{\rm LSND} \sim \frac{\lm^r}{\es} \sim 3 \times 10^{-2}
\,.
\end{equation}
Since this mixing angle is small,
the masses of $\nu_3$ and $\nu_4$  approximately
come from $(22)$, $(23)$, and $(33)$ 
components of $\M_{\nu}$
which implies
the mass difference $\Delta m_{34} \sim \lm^{a}$.
The Super-Kamiokande data are explained as
\begin{equation}
\label{Super-K}
\Delta^2 m_{34} \sim \lm^{a} \left( \frac{m^2}{\lm M}\right)^2
\sim 10^{-3}\,{\rm eV}^2,~~~{\rm for}~\, a=3
\,,
\end{equation}
where (\ref{LSND}) has been used.
Secondly, the masses of $\nu_1$ and $\nu_2$  approximately
come from $(11)$, $(14)$,
and $(44)$ components of $\M_{\nu}$.
Since $t<a$ in our charge assignment, we naturally have
maximal mixing between $\nu_1$ and $\nu_2$
with the mass squared difference as
\begin{equation}
\label{VO}
\sin\theta_{12} \simeq \frac{1}{\sqrt{2}}
~~\& ~~
\Delta m^2_{12} \sim
10^{-10} {\rm eV}^2\,,
\quad {\rm for} ~~\, 
t=1
\,.
\end{equation}
This corresponds to the vacuum oscillation solution for the
solar neutrino problem.

Let us summarize the characteristic features of our mechanism:
\begin{itemize}
\item 
The four neutrino mass spectrum consisting of two almost degenerate pairs
separated by a mass gap $\sim 1$~eV is naturally generated.
\item 
The mass scale of the right-handed neutrinos in
our case is high enough even in the singular seesaw mechanism.  
The understanding of neutrino mass and mixing can be put into the same
category as the charged leptons and quarks.    
\item
The large mixing for the atmosphere neutrino problem is not automatic.  
It is achieved by introducing Abelian family symmetry.  
The $\nu_{\mu}-\nu_{\tau}$ large mixing is
feasible in this case.  
\item 
The lightest neutrino masses are not
generated by seesaw mechanism.  
The large mixing of the vacuum oscillation solution
for the solar neutrino problem can be accommodated.
\end{itemize}

\section{Discussion and Summary}
\label{sec:sd}

In this four light neutrino scenario,
the flavor charge assignment or the texture of the neutrino
mass matrix may imply that the first generation fermions are 
exceptional as far as their masses are concerned.  
For the second and third generations, 
neutrinos have been treated essentially the same as in 
three light neutrino scenarios using the ordinary seesaw 
mechanism\cite{three2}.  
Thus we expect that the charged lepton and quark masses of 
these two generations can be naturally understood within the framework 
of Abelian flavor symmetry.  For the first generation, although it is 
possible to produce appropriate masses of the charged fermions by introducing 
some exotic flavor quantum numbers, the Yukawa couplings 
with flavor symmetry might not be the source for the masses.  In the 
following, we point out a possibility of the mass origin of the first 
generation fermions which was mentioned in Ref. \cite{liu}.  
In the supersymmetric model with R-parity violation and baryon number 
conservation, the following interactions are allowed by the gauge symmetry
and the flavor symmetry,
with positive definite integers $m$ and $n$,  
\begin{equation}
L L E^c \left(
\frac{ \langle X \rangle_{\rm VEV} } {\Lambda}
\right)^m
\,,
\quad
Q L D^c \left(
\frac{ \langle X \rangle_{\rm VEV} } {\Lambda}
\right)^n
\,,
\end{equation} 
where we have suppressed the generation indices.
The $Q$ and $D^c$ are the SU(2) quark doublet and down-type anti-quark singlet 
superfields, respectively.  
We note that if the sneutrino fields get 
non-vanishing VEVs, the above interactions generate masses for the charged 
leptons and down-type quarks.  When the sneutrino VEVs are around 
$\sim \left( \, 10^{-3}- 10\,  \right)$ GeV, the correct magnitudes 
for electron and down quark masses can be obtained\footnote{
A GeV sneutrino VEV can result in a
MeV $\tau$ neutrino mass at tree level in general.  However this conclusion 
is model-dependent.  For example, the model described in Ref. \cite{liu2} 
has vanishing $\tau$ neutrino mass at tree level.}.
One interesting point 
is that the up quark remains massless, 
so that there is no strong CP problem\cite{CP}.  
Recently a similar idea was carried out in detail in Ref. \cite{kim}. 

In summary, within the framework of singular seesaw mechanism, we have
studied the four light neutrino scenario which explains the atmosphere,
solar, and LSND neutrino experiments.  By taking the Dirac neutrino mass
matrix to be also singular, the right-handed neutrino Majorana mass can
be at $\sim 10^{14}$ GeV scale.  Abelian flavor symmetry is used to produce 
the required neutrino mass pattern. 

\acknowledgments

We would like to thank E.J. Chun, C. Giunti, S.K. Kang and C.W. Kim for
various helpful discussions.  J.S. is supported in part by KOSEF
through CTP, SNU.


\end{document}